# Revealing the Mechanism of Electrochemical Lithiation of Carbon Nanotube Fibers

Nicola Boaretto[1,2,3,*], Moumita Rana[2], Rebeca Marcilla[1], Juan José Vilatela[2,*]

1 Electrochemical Processes Unit, IMDEA Energy, Avda. Ramón de la Sagra, 3, 28937 Móstoles, Madrid, Spain

2 IMDEA Materials, Eric Kandel 2, 28906 Getafe, Madrid, Spain

3 Centre for Cooperative Research on Alternative Energies (CIC energiGUNE), Basque Research and Technology Alliance (BRTA), Alava Technology Park, Albert Einstein 48, 01510 Vitoria-Gasteiz, Spain

Corresponding authors

Nicola Boaretto

Email: nboaretto@cicenergigune.com

Phone: (+34) 945 29 71 08

Juan Jose Vilatela

Email: juanjose.vilatela@imdea.org

Phone: (+34) 915 49 34 22

## Abstract

Fabrics of continuous fibers of carbon nanotubes (CNTFs) are attractive materials for multifunctional energy storage devices, either as current collector, or as active material. Despite a similar chemical composition, lithiation/delithiation in CNTFs is substantially different from traditional graphite electrodes. In CNTFs this process is dominated by surface processes, insertion in the bundles interstices, electrochemical doping and often-overlooked partial degradation of the $sp^2$ lattice upon cycling. Through extensive electrochemical analysis, together with in situ Raman spectroscopy measurements, we analyzed the complex lithiation behavior of highly crystalline fibers of CNTs. CNTF can store lithium reversibly with high specific capacity and rate capability, thanks to a large capacitive contribution. Upon lithiation, they undergo electrochemical doping, with longitudinal conductivity increasing by as much as 100 %, concomitant with large downshifts in Raman spectra. However, CNTF are also affected by high first-cycle irreversible capacity, voltage hysteresis and amorphization upon cycling. Electrochemical analysis confirms that SEI formation is responsible for the first-cycle irreversible capacity. Voltage hysteresis is attributed primarily to the trapping of lithium ions in the interstices between stacked nanotubes. Another dominant feature is pre-existing defects, which promote capacitive storage but lead to progressive amorphization of the CNTFs. Indeed, it is evidenced that undesired amorphization is hindered in ultra-pure CNTF without pre-existing defects.

**Introduction**

Macroscopic carbon nanotube fibers (CNTF) are ideal materials for multiple electrochemical applications due to their unusual combination of high electronic conductivity and high toughness.[1] Unlike regular CNT powders, CNTFs can be spun continuously even up to a kilometer. Such features result in high conductivity and mechanical stability in macroscopic scale, making them promising as free-standing current-collector for electrochemical applications. For example, CNTF have been successfully applied as current collectors in lithium-ion batteries,[2–4] lithium-metal batteries[5] and lithium-ion capacitors,[6] including different textile-like architectures such as for wearable applications.[7,8] In addition, CNTF have been used directly as an active material for lithium ions storage. Carbon nanotubes (CNTs) are characterized by high theoretical specific capacity,[9] with reported maximum lithiation stoichiometry of $LiC_3$ and $LiC_2$ for etched and ball-milled tubes, respectively,[10,11] and by higher rate capability than graphite.[12] Thus, the use of CNTF acting as both current collector and active material in Li-ion batteries could result in an increase of both energy and power densities. However, as active material CNTs usually lead to high first-cycle irreversible capacity and do not exhibit a proper plateau during delithiation.[13]

In spite of significant research on the topic, the lithiation of CNTs remains a complex and fascinating subject, first of all because it follows a different mechanism with respect to lithiation of graphite, with no definite staging mechanism.[14–16] The proposed mechanism of CNT lithiation involves insertion of lithium ions in the interstitial space between tubes, and intercalation between the pseudo-graphitic planes inside the tubes. The latter implies diffusion through the tubes open ends (where possible, for example in etched CNTs) and through topological defects.[17,18] In addition, it is widely accepted that the interaction between lithium and carbon atoms has mostly ionic character (i.e. almost complete charge transfer occurs after interaction of lithium atoms with neutral carbon nanotubes),[19–22] and that lithium insertion and intercalation occurs at potentials close to the intercalation of graphite.[17,23,24]

However, fundamental aspects of the lithiation processes in CNTs still remain controversial and unclear. For instance, the origin of the large irreversible capacity has been ascribed alternatively to irreversible lithium storage in the CNT bundles,[25] or to SEI formation.[12,18,26–28] Also the lithiation/delithiation hysteresis has been attributed to a variety of factors, such as the interaction of lithium ions with interstitial carbon atoms or hydrogen[26], electrostatic interaction with surface C-O functional groups,[27] pseudocapacitive effects involving mesoporosity,[13,28] or to the adsorption of lithium ions on active sites, which requires then higher driving force for the desorption process.[29] This relates also to the question of whether CNTs can be regarded as a pseudocapacitive material,[13] and in general about the role of impurities and defects.[21]

All these elements define a complex picture, complicated by the fact that the electrochemical behavior of CNTs depends strongly on the structural and morphological characteristics of the material under study. Large irreversible capacity and sloping voltage profiles are reported for CNTs ranging from defective multiwalled carbon nanotubes (MWCNT) produced by catalytic chemical vapor deposition[30] to highly conjugated single-walled carbon nanotubes (SWCNT) produced by laser ablation,[31] but which might also have differences in impurities, porosity and defect type.

Another relevant aspect which remains unclear is the possible formation of defects upon lithiation, which has been observed as an irreversible increase of the D/G ratio in the Raman spectra of CNTs after lithiation.[18,27,32] To date, it is not clear what is the reason for the nanotubes degradation/amorphization and, consequently, it is difficult to design strategies to mitigate or completely avoid this undesired behavior. As this effect may be relevant also for other possible applications of CNTF, such as current collectors[1] or as highly doped electronic conductors,[33] it is of great interest to understand the electrochemical processes responsible for degradation of CNTs under lithiation, in particular to clarify the role of amorphous carbon, the presence of defects, composition of SEI, etc.

In this report, we analyze the electrochemical behavior of carbon nanotube fibers during electrochemical lithiation/delithiation, using in situ Raman spectroscopy, electrochemical impedance spectroscopy (EIS), longitudinal conductivity measurements and postmortem analysis. We study fibers of few-layer MWCNTs, which have already been used as current collectors in Li-ion battery electrodes.[1,3,34] In order to enable a better understanding of the lithiation of carbon nanotubes in general, we also compare results with highly packed and defect-free DWCNTs fibers.

**Experimental Section**

Fibers of CNTs of few layers[3] (MWCNTF) were synthesized by floating catalyst chemical vapor deposition,[35] using ferrocene and thiophene as catalyst and catalyst promoter, respectively, and butanol as carbon source. The reaction was carried out in hydrogen atmosphere, with a precursor feed rate of 5 mL h$^{-1}$. CNTF fabrics were produced by direct spinning of the CNT fibers from gas phase, with a winding rate of 5 m min$^{-1}$. The synthetic conditions were adjusted to obtain thin fabrics of about 5 μm thickness, with a density of about 0.3 g cm$^{-3}$.[36] Fibers of double-walled CNTs (DWCNTF) with ultra-high purity were obtained from the commercial supplier Dexmat. These fibers are produced by wet-spinning from a liquid crystalline dispersion of high-crystallinity predominantly double-walled carbon nanotubes.[37]

Scanning electron microscopy (SEM) images were obtained using a Helios Nanolab 600i FIB-FEGSEM dual-beam microscope, at 5 kV. Transmission electron microscopy images (TEM) were acquired with a Talos F200X FEG microscope. Thermogravimetric analysis (TGA) was performed with a Q50 thermobalance from TA Instruments, at a rate of 10 °C min$^{-1}$, in air atmosphere.

Voltammetric and impedance tests were performed with a VMP3 Biologic potentiostat, whereas galvanostatic cycling was performed with a Neware BTS 4000 battery tester. Cyclic voltammetries (CV) and galvanostatic charge-discharge were carried out in coin cells, where CNTF fabrics were used as positive electrode and lithium metal foil (Sigma Aldrich) as negative electrode. A 1 M solution of LiPF$_6$ in ethylene carbonate/diethyl carbonate (1:1 v/v, Solvionic) was used as electrolyte. Whatman GF/D and Celgard 2400 discs were used as separators. The latter were placed onto the CNTF to avoid deposition of glass fiber on the CNTF (which would affect the post mortem microscopic analysis). Cell assembly was performed in a glove box, under argon atmosphere, with oxygen and moisture levels below 0.5 ppm.

Cyclic voltammetries (CV) were carried out on the CNTF/Li cells between 0 and 3 V vs. Li/Li$^+$, with scan rates of 0.1, 0.25, 0.5, 1, 2.5, 5, 10, 25, 50 and 100 mV s$^{-1}$, with 5 cycles at each scan rate. In the rate capability tests, CNTF/Li coin cells were charged and discharged galvanostatically between 3 V and 0.01 V (vs. Li/Li$^+$) at the following specific currents: 0.025, 0.05, 0.1, 0.25, 0.5, 1, 2.5, 5, 10 A g$^{-1}$, based on the mass of the CNTF electrode, with 5 cycles at each specific current. After the rate capability test, the cells were charged and discharged 100 times between 3 V and 0.01 V at 0.1 A g$^{-1}$. The mass of the MWCNTF electrodes per area was of about 0.15 mg cm$^{-2}$.

The impedance test was performed in a Swagelok three-electrode cell, with CNTF fabric as working electrode and two lithium metal discs as counter and reference electrodes. The cell was cycled between 3 V and 0 V, by potential step voltammetry, with potential steps of 200 mV, with each step during 5 h. Impedance spectra were acquired at each potential step, at the end of the equilibration time (5 h). The AC amplitude was set to 10 mV, the frequency range from 1 MHz to 10 mHz, and 10 points per frequency decade were recorded. Fitting of the spectra was carried out with the Z fit analysis tool of the EC-Lab software.

The electronic conductivity test was conducted in a glass cell, with a CNTF yarn soaked in liquid electrolyte (the same electrolyte as above), with two pieces of lithium metal as counter and reference electrodes. The two ends of the yarn were connected to copper wires, which were further connected to the working electrode terminals of a first channel of the potentiostat. The counter electrode terminal and the reference electrode terminal were connected to the two pieces of lithium metal, which were also soaked in the electrolyte. A second channel was connected exclusively on the CNTF yarn, the working electrode terminal on one end, the counter and reference terminal on the other end. CV was performed through the first channel, between 3 V and 0 V vs. Li/Li$^+$, at 0.1 mV s$^{-1}$. At regular intervals of potential, the CV experiment was paused and a linear sweep voltammetry was performed through the second channel, at 100 mV s$^{-1}$, in the range ± 50 mV.

Raman spectra of pristine materials and in situ Raman spectra of CNTF electrodes during electrochemical testing were acquired using a Renishaw spectrometer equipped with a 532 nm He−Ne laser. In situ Raman experiments were performed with a two-electrode split test cell with quartz window (EQ-STC-RAMAN, MTI Corp.) with CNTF fabric and lithium metal as positive and negative electrodes, respectively. LiPF$_6$ (1 M) in EC/DEC (1:1 v/v) was employed as electrolyte, and two Whatman GF/D discs were used as separators. The cell was galvanostatically cycled once at 100 mA g$^{-1}$ (based on the CNTF mass) between the open circuit potential (OCP, ca. 2.5 V vs. Li/Li$^+$) and 0.05 V with a Biologic SP200 potentiostat, and the Raman spectra were collected at intervals of 500 mV.

## Results and Discussion

The material used in this study (MWCNTF) is composed of fabrics of CNTFs. As shown in the SEM micrograph in Figure 1a, CNTFs consist of a porous network of long CNT bundles. The latter have average diameters of about 50 nm and represent domains over which the CNTs are aligned and in close contact (Figure 1b), analogous to elongated crystals in turbostratic graphite. At higher magnification, typical TEM micrographs in Figure 1c and Figure S1 show that the individual CNTs have few layers (3-5) and are highly regular. Indeed, the Raman spectrum is indicative of a highly graphitic material with a D/G intensity ratio of $0.18 \pm 0.07$, and an asymmetric G peak (1582 cm$^{-1}$) combining a shoulder at higher wavenumbers corresponding to the D' (Figure 1d) component. The surface area and pore volume of the CNTF are of ~250 m$^2$ g$^{-1}$ and 0.89 cm$^3$ g$^{-1}$, respectively.[36] As in most CNTs produced by chemical vapor deposition, in addition to CNTs the material contains amorphous carbon and residual metallic catalyst, which amount to ca. 9 wt% and 5 wt%, respectively (Figure S2). These CNTs have a length ranging from 100 μm to 1 mm and have at least one end closed by the catalyst particle.[36]

**Cyclic voltammetry**

Electrochemical lithiation of MWCNTF was first investigated by cyclic voltammetry (CV) in MWCNTF/Li coin cells. Figure 2a shows cyclic voltammograms at 0.1 mV s$^{-1}$. During the first cathodic scan (3 V → 0 V vs. Li/Li$^+$), a broad and weak peak is observed at about 1.5 V (Ic), followed by an intense peak at 0.7 V (IIc). Below 0.5 V, the cathodic current increases monotonically till the vertex potential is reached (IIIc). In the corresponding anodic scan, four weak peaks are observed at 0.2 V, 1.2 V, 2.0 V and 2.3 V vs. Li/Li$^+$ (Ia, IIa, IIIa and IVa, Figure 2b). In the second cycle (Figure 2a), the cathodic peaks at 1.5 V (Ic) and at 0.7 V (IIc) have much lower intensity than in the first cycle. Moreover, the intensity of peak Ia and separation of peaks IIIa and IVa increases slightly (Figure 2b). The spectra feature remains essentially unchanged in the following cycles.

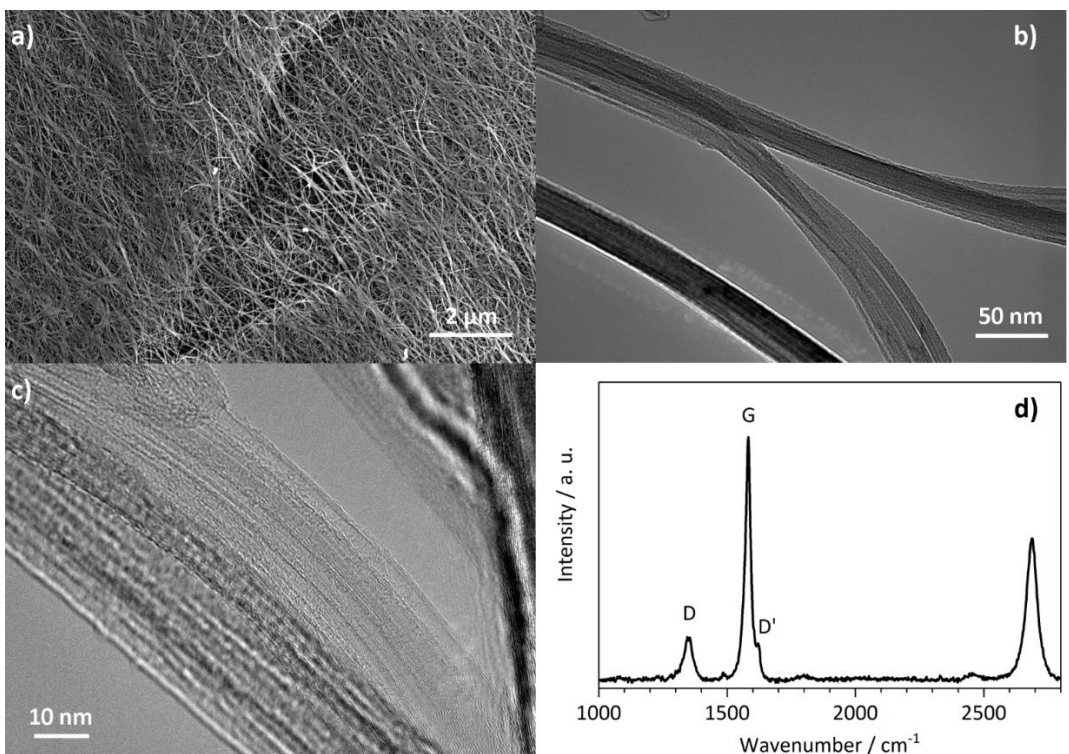

***Figure 1.*** *a) SEM image; b), c) TEM images; d) Raman spectrum of pristine CNTF fabrics (MWCNTF).*

The two peaks Ic and IIc are assigned to irreversible processes, as their intensity decreases already in the second cycle. The peak Ic is tentatively attributed to reduction of impurities, traces of humidity, or of surface oxygenated functional groups. The intense cathodic peak at 0.7 V (IIc) is usually attributed to SEI formation[26] and the high current associated to this process is in good agreement with the high surface area and porosity of CNTs in comparison with graphite. This peak is observed (although with much lower intensity) also in the following cycles (Figure 2b), indicating that the SEI formation is not complete after the first cycle. Finally, the strong current increase below 0.5 V (IIIc) is attributed to the reversible lithiation process. In the anodic scan, the weak anodic peak at 0.1 V (Ia) is attributed to a fast delithiation process. As noted previously, the intensity of Ia increases during the following cycles (Figure 2b). This suggests that this peak may be due to delithiation on defective regions created during the first cycle. The area under peak Ia does not match the one of the corresponding lithiation peak, thus indicating that delithiation occurs partly at higher potentials, as usually observed in CNTs.[13,23] The peaks IIa and IIIa are also attributed to reversible delithiation, as their intensity does not decrease upon cycling. The displacement of the two delithiation peaks to higher potentials confirms the presence of hysteresis. As discussed in the introduction, hysteresis is due to trapping of lithium ions in cavities or to interaction with defects or functional groups.[26–29]

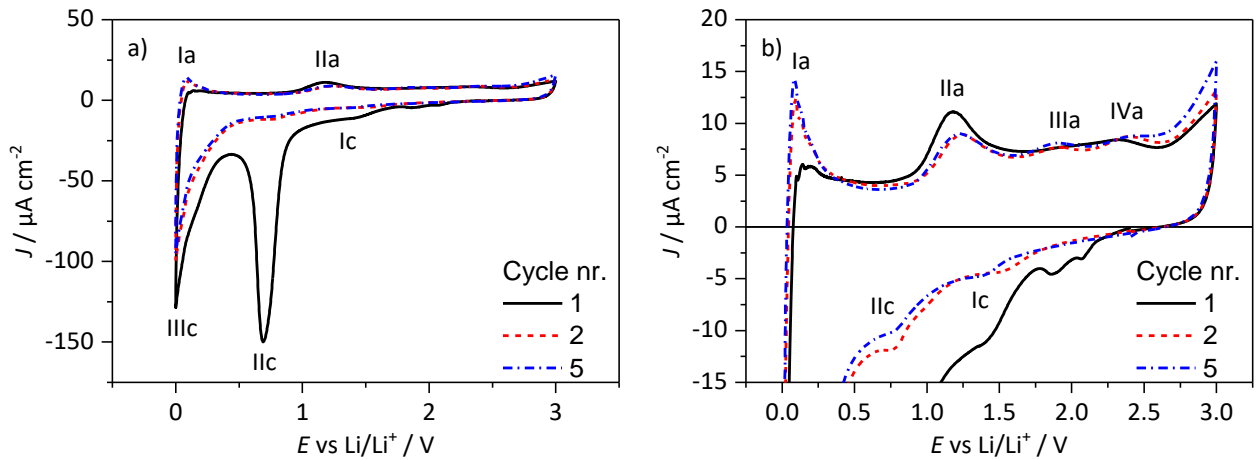

**Figure 2.** a) Cyclic voltammograms of MWCNTF vs Li at 0.1 mV s$^{-1}$, first, second, and fifth cycles. b) Zoom of the cyclic voltammograms at 0.1 mV s$^{-1}$.

To determine whether the lithiation/delithiation of MWCNTFs has predominantly diffusive or capacitive character, the dependence of the current ($i$) on scan rate ($v$) was analyzed. The CVs were carried out at various scan rates, between 0.1 and 100 mV s$^{-1}$ (Figure 3a). At a fixed potential, the current is expected to follow a power dependence on the scan rate:[38]

$$i(V) = av^b \qquad (1)$$

where $a$ and $b$ are adjustable parameters. For diffusion-limited redox reactions, the current is proportional to the square root of the scan rate, thus $b$ = 0.5. On the other hand, capacitive current is proportional to the scan rate, i.e. $b$ = 1. When both capacitive and diffusional effects are present, $b$ is expected to be comprised between these limits. We analyzed the intensity of peaks IIIc, Ia and IIa as a function of the scan rate. The peak intensities follow the power law up to up to 1 mV s$^{-1}$, i.e. log$i \propto$ log$v$ (Figure 3b). Within this threshold, potential drops due to electrolyte resistance and peak potential displacements are negligible. The power law exponent $b$ was then calculated from the slope of the double-logarithmic plot. $b$ is 0.99 and 0.97 for Ia and IIa respectively, indicating main capacitive character of these processes. Conversely, $b$ = 0.67 for peak IIIc, suggesting a mixed diffusive/capacitive control. In the next step, we quantified the diffusive and capacitive contributions at fixed potentials using the following formula:[39]

$$i = k_1 v + k_2 v^{1/2} \tag{2}$$

where $k_1 v$ and $k_2 v^{1/2}$ are the surface-controlled (capacitive) and diffusion-controlled current contributions, respectively. The analysis was carried out in the scan rate range between 0.1 and 1 mV s$^{-1}$, every 100 mV. The resulting sets of coefficients $k_i$ were then used to reconstruct the CVs. The experimental and reconstructed CV at 0.1 mV s$^{-1}$ are reported in Figure 3c, with the area in red representing the capacitive contribution associated to the first term in equation 1. It is noteworthy that even at low scan rates the capacity term is predominating (68 % at 0.1 mV s$^{-1}$, Figure 3d). Significant diffusive contribution is observed only during the cathodic scan, below 0.5 V, and in the anodic scan over 2 V. At higher scan rates the capacitive contribution obviously increases, reaching over 80 % at 1 mV s$^{-1}$ Figure 3d). The results indicate that lithiation below 0.5 V vs. Li/Li$^+$ occurs under mixed capacitive/diffusion limitation. Delithiation, even if affected by hysteresis, occurs predominantly as a capacitive process.

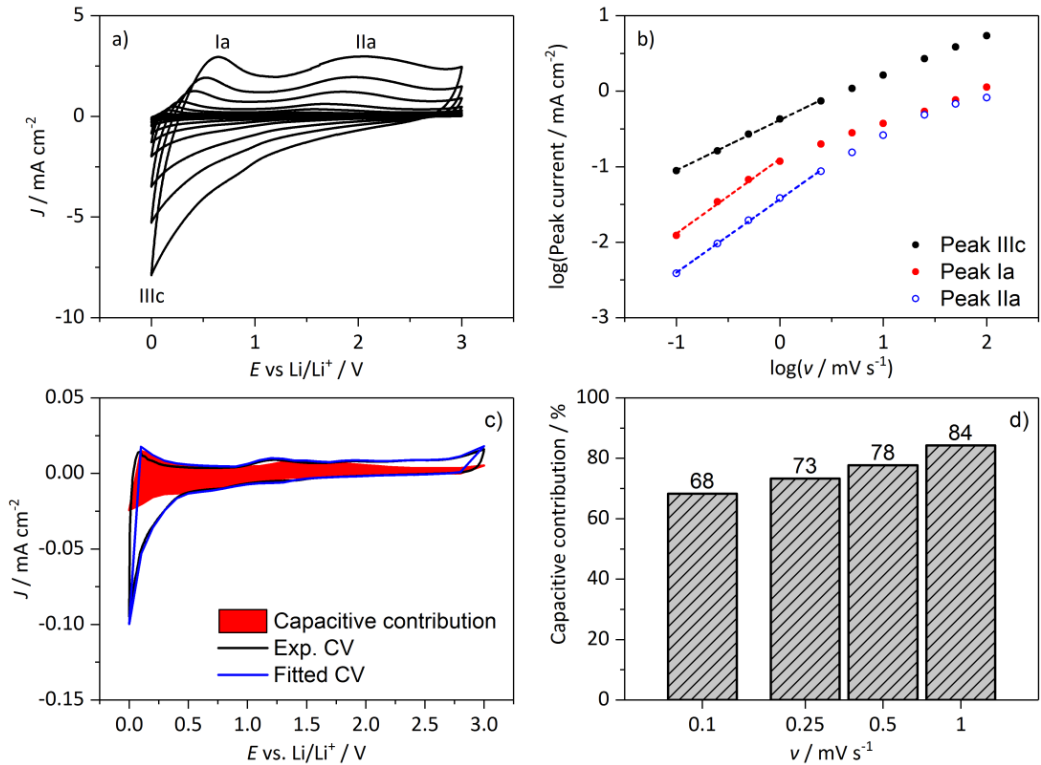

*Figure 3. a) CVs of MWCNTF at scan rates between 0.1 and 100 mV s$^{-1}$; b) Log-log plot of the intensities of peaks IIIc, Ia and IIa versus the scan rate; c) Experimental CV at 0.1 mV s$^{-1}$, the same CV reconstructed with equation 1 (in blue), and the correspondent capacitive term (in red); d) capacitive contributions at different scan rates from 0.1 to 1 mV s$^{-1}$.*

For comparison, cyclic voltammetry was performed also with commercial DWCNTF (see experimental section). With respect to in-house MWCNTF, these DWCNTF are open-ended and have very low concentration of preexisting defects. The CVs show similar patterns (Figure S3); thus confirming the presence of hysteresis and SEI formation even in defect-free DWCNTF. SEI formation is even more pronounced than in MWCNTF, indicating a larger contact area with the electrolyte. Interestingly, in DWCNTF the anodic peak Ia is observed only in the second and following cycles, with intensity increasing upon cycling. The intensity of this peak is in any case lower than in the CVs of MWCNTF (ca. one third by normalizing on the intensity of IIIc). The absence of peak Ia in the first cycle is attributed to the absence of defects in the pristine DWCNTF. The low intensity of the same peak in the following cycles suggests that defect formation is hindered in DWCNTF.

**Galvanostatic cycling**

The electrochemical behavior of MWCNTF/Li was further studied by galvanostatic cycling. The potential-capacity curve during the first lithiation (Figure 4a) displays large steady plateau at ~1.0 V and a sloped plateau below 0.5 V. In agreement with CV analysis, the first plateau is attributed to SEI formation and the second one to reversible CNTF lithiation. The reversible capacity is ca. 0.4 Ah $g^{-1}$, i.e. only 30 % of the capacity reached during first lithiation (1.4 Ah $g^{-1}$). The reversible capacity is thus close to that of graphite (372 mAh $g^{-1}$) but the shape of the voltage profile appears to be quite different. As generally observed for CNTs,[14–16] no defined plateau is observed during delithiation (Figure 4a). After the first cycle, the potential decrease during lithiation resembles an exponential decay, flattening towards the cutoff voltage and trespassing smoothly to the lithium plating region at 0 V.

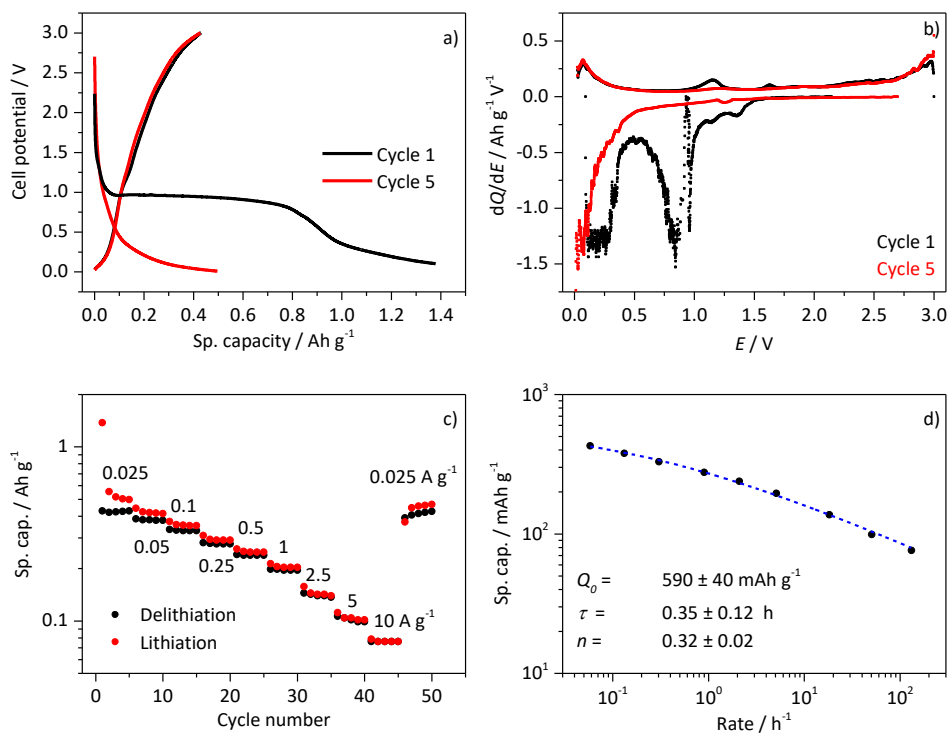

**Figure 4.** *Galvanostatic cycling of MWCNTF/Li cells: a) potential-capacity profiles in the first and fifth cycle (constant current of 25 mA g$^{-1}$); b) Differential capacity profiles of the first and fifth cycle; c) Specific capacity at various current densities; d) Double-logarithmic plot of the specific capacity versus the effective rate (see equation 3).*

The delithiation curves are more complex and show three distinctive regions: below 1 V the shape of the potential profile resembles the one of the lithiation profile, corresponding to the "fast" delithiation process (peak Ia) observed in the CVs (Figure 2a). Between 1.0 V and 2.5 V, the potential increases more slowly, approximately linearly with the capacity. This region corresponds to the hysteresis-affected delithiation. At higher potentials, a pseudo-plateau marks the onset of the "overcharging" region, in which irreversible processes, such as reoxidation of SEI and partially irreversible charge transfer at preformed defects might take place.

The differential capacity profiles (Figure 4b) show features similar to the CVs: in the first lithiation, two intense peaks are attributed to the SEI formation (~ 1.0 V), and reversible lithiation of (below 0.5 V) whereas in the first delithiation fast and slow delithiation components are observed at potentials < 0.5V and > 1.0V, respectively. The profiles in the following cycles, conversely, show mainly the peaks corresponding to reversible lithiation and to the fast delithiation processes both below 0.5 V.

Galvanostatic cycling between 3.0 V and 0.01 V was performed at various current densities to study the rate capability of CNTF as anode material (Figure 4c). At 0.025 A g$^{-1}$ MWCNTF deliver a capacity of about 428 mAh g$^{-1}$ whereas at current densities as high as 10 A g$^{-1}$ (5 mA cm$^{-2}$), CNTF deliver a reversible specific capacity of about 76 mAh g$^{-1}$. This corresponds to a capacity retention of 18 %. Following Tian et al.,[40] the dependence of the capacity Q with the current can be modeled through:

$$Q = Q_0[1 - (R\tau)^n(1 - e^{-(R\tau)^{-n}})] \quad (3)$$

where Q is the capacity, $Q_0$ is the zero-rate limiting capacity, R is the effective rate, i.e. the ratio between capacity and current, $\tau$ is a time constant and n indicates the type of rate-limiting behavior (n = 0.5 for diffusion-control and n = 1 for capacitive-control processes). The specific capacity as a function of the rate and the correspondent fitting parameters are reported in Figure 4d. The value of $Q_0$ is close to 0.6 Ah g$^{-1}$, 50 % higher than the capacity registered at 25 mA g$^{-1}$. This limiting capacity corresponds to a stoichiometry close to LiC$_4$, which most likely involves lithium insertion inside the CNTs and interaction with the nanotube inner layer. However, the analysis indicates that such limiting capacity is reached only at extremely high times scales (> 10$^4$ h) and at practical current densities the surface capacitive contribution predominates. The low value of the exponent n = 0.32 indicates a limited dependence of the capacity with the rate R, in the examined range of rate. At even higher rates, the capacity is expected to drop faster due to mass transport limitations, which would result in an increase of n to, at least, 0.5.

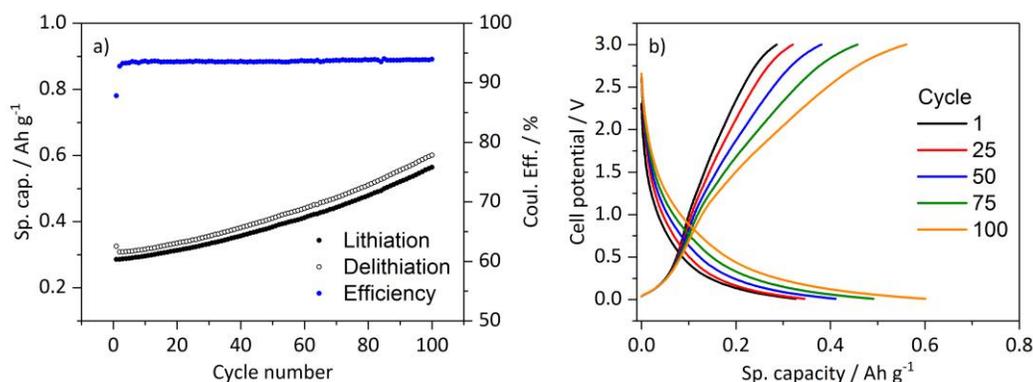

*Figure 5. Cyclability test of MWCNTF/Li cells (0.1 A g$^{-1}$): a) specific capacity and coulombic efficiency; b) potential/capacity profiles.*

After the rate capability test, MWCNTF/Li cells were further cycled for 100 cycles at a constant current of 0.1 A g$^{-1}$ (Figure 5a). Interestingly, the capacity steadily increases from 0.3 Ah g$^{-1}$ up to ca. 0.6 Ah g$^{-1}$ after 100 cycles. Meanwhile, the average coulombic efficiency is about 94 % during the whole cycling period, indicating that parasitic reactions occur throughout the test. Another cell, cycled more extensively, showed capacity peaking at ca. 0.9 Ah g$^{-1}$ (Figure S4), close to the capacity of soft amorphous carbons.[41] To investigate further the origin of the increasing capacity, we looked into the voltage profiles during the cyclability test (Figure 5b and S5). The lithiation curve maintains the original shape (see normalized capacity in Figure S5) but a decrease in the slope of the potential-capacity curve during delithiation is observed, indicating that hysteresis is aggravated upon cycling. These results suggest that the concentration of surface defects or accessible interstitial sites increases upon cycling. Capacity increase upon cycling with MWCNT was previously observed by Masarapu et al.[42] In that case, the capacity increase was attributed to the combined effect of a) the creation of more active sites for lithium storage due cleavage of the nanotubes upon cycling and, b) increased exposure of the acid-treated stainless-steel current collector, contributing to the electrode capacity through lithium storage on the iron oxide surface film. While we don't expect any significant contribution of the current collector (we used normal coin cell stainless steel plates, with no acid pre-treatment), we do observe degradation of the nanotubes by both Raman spectroscopy and TEM (see discussion below) and, concordantly with the cited study, attribute the increased capacity to the subsequent formation of new defects and Li storage sites.

Another property which is affected by the cycling is the self-discharge behavior. Pristine MWCNTF are characterized by self-discharge of about 1 V, even after maintaining the potential at 10 mV *vs.* Li/Li$^+$ for 100 h (Figure S6). However, after 100 cycles, the self-discharge is greatly reduced and the cell potential relaxes back only up to 0.2 V. This is attributed to an increase of lithiation at surface/subsurface defects in graphitic planes of the CNT fibers, which are characterized by a higher binding energy.[21]

The cycling behavior of the reference, defect-free DWCNTF confirms and complements the assignation of the processes discussed above. The capacity underlying the irreversible plateau at 1.0 V (ca. 3 Ah g$^{-1}$) supports the assignment of the high-voltage plateau to SEI formation, as the correspondent stoichiometry is not compatible with any kind of lithium insertion mechanism, unless lithium aggregates with almost metallic character were formed. If this were the case, we expect it to occur close to the lithium plating potential (ca. 0 V). The reversible capacity reaches 700 mAh g$^{-1}$ (Figure S7a), corresponding approximately to a stoichiometry of LiC$_3$ and almost doubling the reversible capacity of high-throughput fibers of few-layer MWCNTFs (ca. 400 mAh g$^{-1}$) (Figure 4a). This high value cannot be attributed to defects due to the defect-free character of the investigated DWCNTF and must therefore include lithium ion accumulation in the nanotube core. It is worth mentioning that such mechanism is facilitated in these samples as the constituent nanotubes have open ends and are significantly shorter (around 1 µm) than the in-house MWCNTF.

The cycling test on the reference material shows that this material also exhibits hysteresis (Figure S7a-c). Since it cannot be attributed to interaction of lithium ions with defects, it must originate from trapping of lithium ions in the interstitial spaces of the bundles or in the nanotube cores. The rate capability is worse than in the MWCNTF, with a capacity of 17 mAh $g^{-1}$ at 10 A $g^{-1}$ (Figure S7d), corresponding to the 2.4 % of that obtained at 25 mA $g^{-1}$. The lower capacity retention may be due to a smaller capacitive contribution due to the lower concentration of defects in this sample. In the cyclability test (at 0.1 A $g^{-1}$), no capacity increase is observed, and the capacity remains constant at about 0.3 Ah $g^{-1}$, with a mean coulombic efficiency of 98.3 % (Figure S8). The absence of capacity increase, together with the results by Raman spectroscopy (see discussion below), indicating that this material is not affected by amorphization upon cycling, further confirms that the capacity increase is due to CNTF degradation.

**In situ Raman spectroscopy**

The analysis of the electrochemical behavior suggests that the defects in MWCNTF play a crucial role in the lithiation mechanism. Taking advantage of the strong resonance of low-diameter CNTs under Raman spectroscopy and the sensitivity of the Raman spectrum to defects, we performed in situ measurement during galvanostatic cycling of MWCNTF /Li cells.

Figure 6a and Figure 6b show the Raman spectra of MWCNTF, between 1200 and 1750 $cm^{-1}$, during the first galvanostatic lithiation and delithiation, respectively. In this restricted range, no signal from the electrolyte is observed, whereas the MWCNTF gives rise to two strong bands at 1350 $cm^{-1}$ and at 1582 $cm^{-1}$, which as noted previously, are attributed to the D and G bands, respectively (Figure 1d). In the initial state (~2.5V) the ratio of the areas underlying the two peaks D and G ($A_{D/G}$) is 0.25 (Figure 6c). The corresponding intensity ratio of the two peaks is 0.17, in agreement with the typical value of pristine CNTF (see experimental section). During lithiation, $A_{D/G}$ increases progressively, reaching about 0.5 at 1.5 V. Between 1.5 V and 1.0 V, $A_{D/G}$ increases abruptly reaching about 2 at 1.0 V. At lower voltages, both the $A_{D/G}$ and the intensity of the individual D and G peaks decrease until disappearing completely at 0.05 V. During delithiation, $A_{D/G}$ ratio stabilizes at 2.5 in the whole potential range. The increase of $A_{D/G}$ is a clear indication of the formation of lithiation-driven stable defects, as already observed in previous studies.[18,26,32,43] These may be caused by the partial covalent interaction between lithium ions and functional groups (e.g. C-O or C-H) on edges or other preexistent defective regions. Mixed ionic-covalent Li-C interaction at high lithiation degrees was already proposed by Claye et al.[32] This partial covalent interaction would result in the propagation of the defects and partial amorphization of the CNT bundles. Since the defects are stable, it is probable that the lithium ions involved in the defect formation are irreversibly bound to the CNTF. The loss of intensity at low potential during lithiation, on the other hand, is due to the loss of resonance condition, and is usually observed at high lithiation degrees both in graphite and CNT.[44]

Besides the peak intensity, also the peaks positions change significantly upon lithiation (Figure 6d). A strong downshift is observed for the D peak below 1.0 V (-22 cm$^{-1}$ at 0.25 V), whereas the G peak is shifted upwards (+15 cm$^{-1}$ at 0.25 V). The marked downshift of the D band indicates that lithium interacts strongly with defects causing, in this case, a decrease in the C-C bond strength. The upshift of the G band corresponds to the formation of diluted graphite intercalation compounds.[45] The upshift of the G peak is accompanied by a broadening and loss of symmetry with a marked tail at low wavenumbers at low voltages (Figure S9a and Figure S9b). This asymmetry corresponds to a Breit-Wigner-Fano interference, which is observed in nanotubes and graphite intercalation compounds at high doping levels. The resulting broad band was fitted by three lorentzian peaks centered at 1565, 1590 and 1610 cm$^{-1}$ (Figure S9b), assigned to the $E_{2g}$ mode of carbon atoms interacting with lithium ions, to the $E_{2g}$ mode of non-interacting carbon atoms, and to the down-shifted D* mode, respectively.

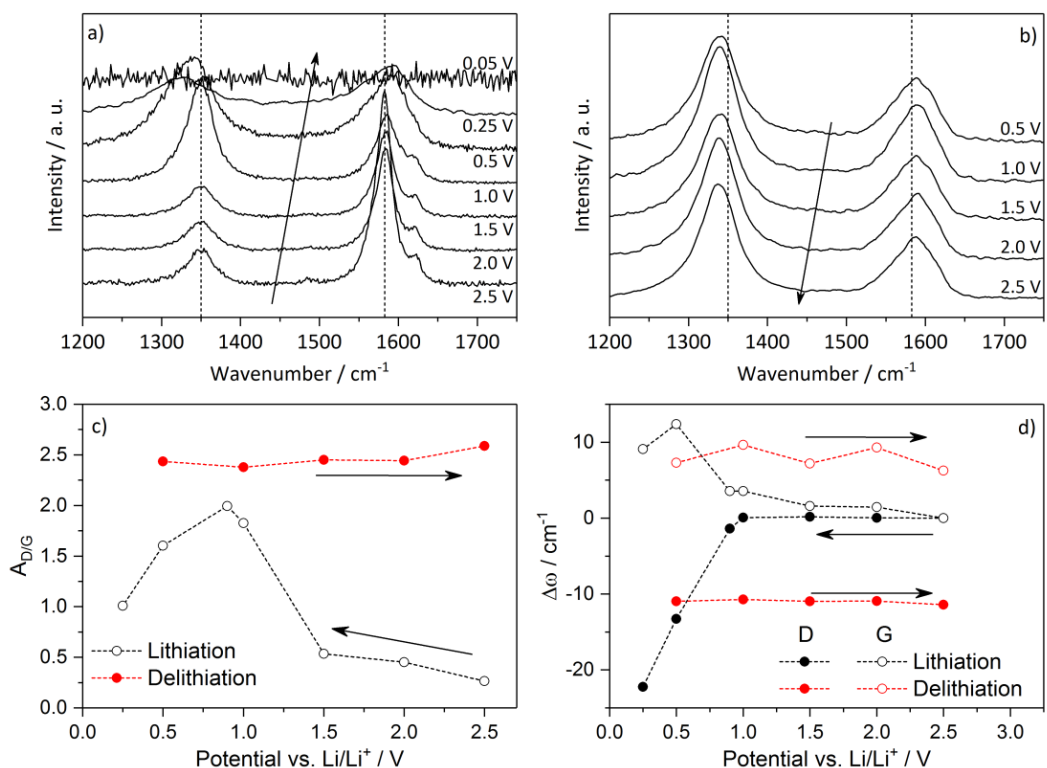

*Figure 6. Raman spectra of pristine MWCNTF during the first galvanostatic cycle, a) during lithiation, and b) during delithiation. The dotted lines in Figure 6a and 6b indicate the initial position of the peaks D and G, in the spectra of the pristine MWCNTF, added to highlight the shift of the two bands with the cell potential. c) Ratio of the areas of peaks D and G as a function of the cell potential; d) Position of the D and G bands as a function of the cell potential.*

Ex situ Raman spectra were also acquired on the reference defect-free DWCNTF (Figure S10). The Raman spectrum of the pristine sample shows a very weak D peak, confirming their defect-free nature. Differently from MWCNTF, Raman spectrum of DWCNTF did not change after the first lithiation and only after 50 galvanostatic cycles the D/G intensity ratio slightly increases to 0.2 (note that this is close to the initial value at OCV for MWCNTF). This confirms that amorphization is strongly dependent on the density of pre-existing defects and can be drastically mitigated by using defect-free materials.

To summarize, our observations from the in situ and ex situ Raman spectroscopy experiments corroborate that in parallel with the gradual amorphization of CNTs, Li ions intercalated lead to strong electrochemical doping, similarly to graphite intercalation compounds.

**Morphological analysis of cycled CNTF**

The effect of electrochemical cycling on the morphology of MWCNTF was analyzed by SEM and TEM. SEM images obtained on galvanostatically cycled samples (100 cycles at 0.1 A g$^{-1}$, Figure 5) show the formation of a SEI as a dense amorphous deposit of insoluble electrolyte decomposition products (Figure 7a). SEM micrographs taken on a broken sample confirm that the CNTs remain very long, with no indication of that amorphization leads to significant shortening of CNTs (Figure 7b). TEM images were acquired on a CNTF sample which had been used only for cyclic voltammetry (Figure 7c-d) in order to facilitate discrimination of the CNTs over the SEI. TEM images shows that the CNTs undergo mainly surface amorphization, particularly on outer layers and in the perimeter of CNT bundles, keeping the inner graphitic backbone preserved. The implication is that amorphization concentrates initially on more exposed graphitic shells and then propagates inside the bundles and inside inner layers of the CNTs.

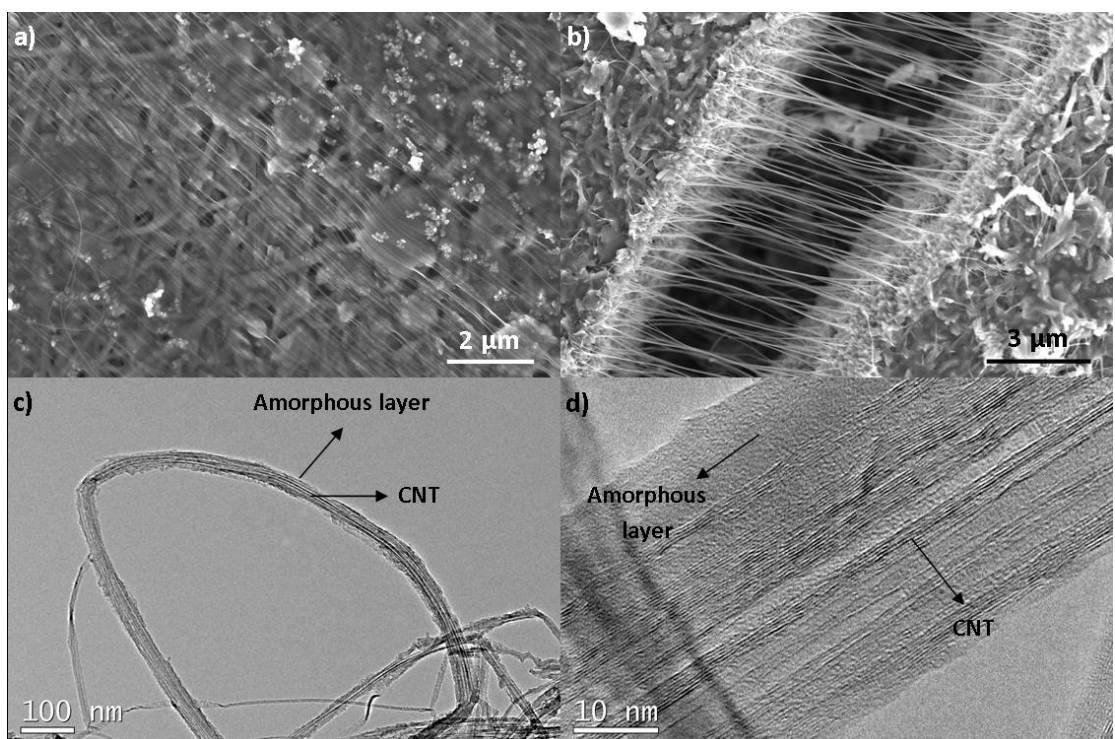

*Figure 7*. SEM and TEM images of cycled MWCNTF. a) SEM image of the horizontal section of a MWCNTF electrode after 100 galvanostatic cycles and b) SEM image of a broken sample of cycled MWCNTF; (c, d) TEM images of MWCNTF after 50 CV cycles.

**In-situ impedance spectroscopy and electronic conductivity**

Impedance spectroscopy measurements in three-electrode configuration were performed on MWCNTF-Li cells at intervals of 200 mV, during one charge and discharge cycle (3 V – 0 V – 3 V vs. Li/Li$^+$). In the pristine state, the spectrum of the MWCNTF working electrode is composed of a skewed and depressed semicircle, followed at low frequencies by a tilted straight line (Figure 8a). The spectrum was modeled by the equivalent circuit shown in Figure S11a comprising: i) a high frequency inductance $L_1$, followed by a series resistance $R_b$, corresponding to the sum of all purely resistive contributions, including the electronic contact resistances and ionic resistance of the electrolyte; ii) a Gerischer element (G), with impedance of the form $Z(\omega) = Z_0/\sqrt{k + j\omega}$, accounting for the skewed semicircle;[46] and iii) a low frequency capacitive tail $Q_{LF}$, assigned to the bulk chemical capacitance of the MWCNTF electrode. The Gerischer-type impedance is represented by a semi-infinite transmission line (Figure S11b) with in each unit length a series resistance ($R_s$), and a parallel combination of a capacitance ($C_p$), and a resistor ($R_p$). $R_s$ represents the ionic transport resistance, $R_p$ a charge transfer resistance and $C_p$ a double layer capacitance. This type of impedance is typical for systems in which diffusion and chemical reactions occur simultaneously, and could thus represent the simultaneous occurrence of lithium diffusion in the CNT bundles and charge transfer process at the bundles surface. Two parameters can be extracted experimentally, namely the distance between the two intercepts of the skewed semicircle $R_G$, and the rate constant $k = \tau_G^{-1}$. $R_G$ is a combination of the two resistive components, whereas $k$ is the rate constant of the chemical reaction. Obviously, the skewed semicircle can be modeled also by multiple RC circuits. However, fitting with two or more RC circuits results in a high uncertainty for the fitting parameters. Using a Gerischer element, on the other hand, results in a good fit with only the two parameters $R_G$ and $\tau_G$.

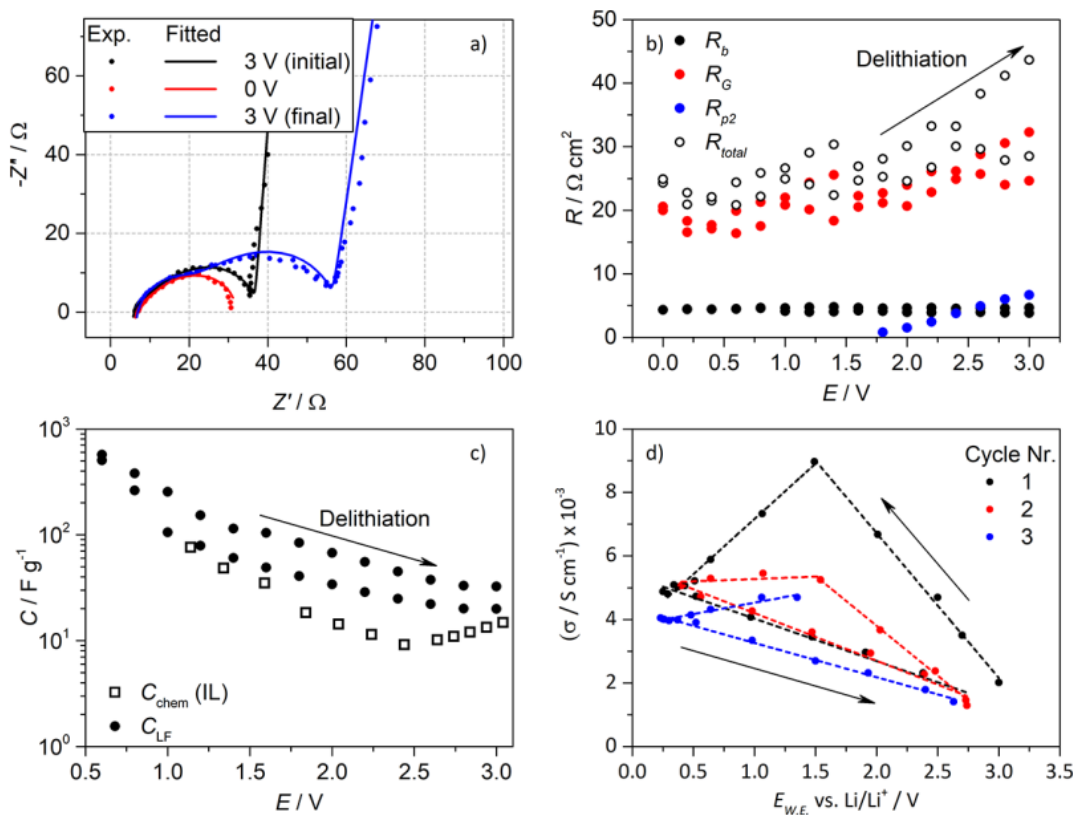

***Figure 8.*** *(a) Impedance spectra of the pristine MWCNTF electrode (black trace), after the first lithiation (red trace), and after the following delithiation (blue trace). The markers represent the experimental data and the lines the fitted spectra. (b) Bulk resistance ($R_b$), interface resistance components ($R_G$ and $R_{p2}$), and total resistance ($R_{total}$) at each potential step of the lithiation/delithiation cycle. (c) Bulk capacitance ($C_{LF}$) vs the potential. The open symbols in (c) ($C_{chem}$) represent the chemical (quantum) capacitance of CNTF fabrics in ionic liquid media (data taken from ref 36). (d) Electronic conductivity of a CNTF yarn during three lithiation/delithiation cycles by CV.*

Analysis of impedance spectra during lithiation and delithiation show the convolution of strong doping, charge accumulation and partial amorphization, confirming observations by Raman and electrochemical measurements discussed above (impedance spectra and the most important fitting parameters are included in Figure S12 and Table S1).

With decreasing potential, $R_G$ drops from the initial 25 to 15 Ω cm$^2$ at 0.5 V (Figure 8b). This is attributed to a decrease of the charge transfer resistance upon doping of the CNT-based material.[14] Below 0.5 V, $R_G$ increases slightly, possibly due to SEI formation. It is noteworthy that no additional spectral features are observed at low potentials, suggesting formation of a low resistive SEI. During delithiation, from ca. 1.8 V, another semicircle is though observed. From this potential an additional RC circuit is required, characterized by a resistance $R_{p2}$ and a capacitance $C_{p2}$ (Figure S11c). $R_{p2}$ increases with the potential, causing a neat increase of the total resistance. The latter reaches almost 45 Ω cm$^2$ at 3 V, a 50 % increase with respect to the initial value (28 Ω cm$^2$). The additional impedance is attributed to the formation of a delithiated amorphous layer surrounding the MWCNTF bundles.

The rate constant $k$ is initially 1.5 × 10$^3$ s$^{-1}$. $k$ increases slightly during lithiation, up to 2.0 × 10$^3$ s$^{-1}$, and decreases significantly during delithiation, reaching 0.8 × 10$^3$ s$^{-1}$ at 3.0 V (Table S1). Also the decrease of $k$ is attributed to the amorphization process.

Finally, the low frequency tail shortens with decreasing potential and, below 0.6 V, the spectra show only the mid-frequency semicircle, thus resembling the spectra of non-polarizable electrodes. Indeed, at this high doping level the Raman spectrum fades as the MWCNTF material resembles a metal.

The capacitance $C_{LF}$ was calculated from the CPE parameter $Q_{LF}$ as proposed by Brug et al. (see footnotes of Table S1).[47] $C_{LF}$ is close to 20 F g$^{-1}$ before the cycle. Upon lithiation, $C_{LF}$ increases steeply, almost two orders of magnitude between 3 V and 0.5 V. Below this potential, the capacitive tail is not observed. Correspondingly, $C_{LF}$ decreases again during delithiation, reaching 33 F g$^{-1}$ at 3 V. The magnitude and dependence of $C_{LF}$ follow those of the chemical (quantum) capacitance of MWCNTF fabrics upon electrochemical doping in ionic liquid media (Figure 8c),[36] thus confirming the electrochemical doping of CNTF in organic electrolyte. The parallel process of amorphization of the material manifests in an 66 % increase of the bulk capacitance from the initial delithiated state. The increase of $C_{LF}$ is probably related to the increase of capacity observed upon cycling (Figure 5).

The picture that emerges from impedance and Raman spectroscopy is that the MWCNTF is highly doped during the first lithiation. The effects of electrochemical doping on bulk longitudinal conductivity of CNT fibers were measured by performing a CV with the MWCNTF as working electrode, two pieces of Li metal as counter and reference electrodes, and measuring the electronic resistance of the fiber at distinct potentials using a second circuit. As shown in Figure 8d, during lithiation conductivity increases linearly up to 1.5 V, from 2 x 10$^3$ S cm$^{-1}$ to ca. 9 x 10$^3$ S cm$^{-1}$. The conductivity returns to a value close to the initial one upon delithiation. Such increase in electronic conductivity is due to electrochemical doping of the CNT fibers, as previously observed in related systems[14,48,49] and consistent with changes in charge transfer resistance (Figure 8b). However, the relation between longitudinal conductivity and charge transfer resistance is not linear. The latter starts increasing at 0.5 V, whereas the electronic conductivity decreases already below 1.5 V. The increase of the charge transfer resistance is attributed to the SEI formation, whereas the decrease of electronic conductivity is attributed to the doping-prompted amorphization. The conductivity dependence on potential is particularly strong in the first cycle. In spite of the apparently large amorphization of the material, bulk longitudinal conductivity remains very high even after multiple cycles, although it decreases relative to the pristine material due to the introduction of defects.

## Conclusion

This report focuses on the electrochemical behavior of multi-walled carbon nanotube fibers (MWCNTF) upon lithiation, in view of a possible utilization of CNTF fabrics as anode material in lithium batteries.

The specific capacity of MWCNTF is close to that of graphite (ca. 400 mAh g$^{-1}$). Analysis of the capacity at various rates suggests that capacities up to 600 mAh g$^{-1}$ can be reached at very large time scales. This capacity corresponds to a high lithiation degree, similar to that observed with etched CNTs,[10] thus indicating that lithium ions can penetrate inside the tubes, although only at extremely slow rates. The rate capability is excellent, with a reversible capacity of 76 mAh g$^{-1}$ at 10 A g$^{-1}$. However, MWCNTF show also severe drawbacks, such as high irreversible capacity during the first cycle (70 %), mostly due to SEI formation, absence of lithiation plateaus and high voltage hysteresis (ca. 1 V) between lithiation and delithiation. In addition, degradation upon lithiation is observed, i.e. partial loss of crystallinity and formation of stable defects, resulting in a slight decrease of the electronic conductivity. This effect was previously reported for CNTs, but no clear explanation was provided for the mechanism of degradation. We performed thus a detailed study to understand this phenomenon and the lithiation mechanism in general. Results show that degradation occurs already at the early stages of lithiation. Since in this region lithium ions interact predominantly with surface defects, the effect is attributed to the formation of partially covalent bonds between lithium ions and surface functional groups, resulting in the reorganization of the neighbor C-C bonds and in the propagation of defects. Comparison with defect-free double-walled CNTF (DWCNTF) confirms that degradation is prompted by the presence of pre-existing defects. Indeed, degradation is hindered in these DWCNTF. Conversely, hysteresis is observed also in absence of defects. In this case, the phenomenon is attributed to the trapping of lithium ions in the interstices between stacked nanotubes. Nonetheless, the increase of hysteresis upon cycling in defect-containing MWCNTF indicates that interaction with amorphous carbon also contribute to such effect.

With regards to the lithiation mechanism, CNT fibers accumulate charge through a combination of different processes. Lithiation occurs at first in the amorphous domains, with formation of partly covalent bonds, and expansion of the same amorphous domains. This process has capacitive-like kinetics. In a second phase (below 0.5 V), lithium inserts in the CNT bundles, with partly diffusion-limited kinetics. Lithium intercalation in the nanotube cores is hindered due to the nanotubes capping, but it may become possible for longer times upon nanotubes degradation. Lithiation results in electrochemical doping of the nanotubes, confirmed by the simultaneous downshift of Raman bands and increase of the chemical capacitance, and results in a 100 % increase of the longitudinal electronic conductivity.

From the application viewpoint, the study indicates that CNTF fabrics are not suitable as anode material for Lithium-ion batteries, mainly due to the high first-cycle irreversible capacity and to the high voltage hysteresis. On the bright side, CNTF show high capacity, rate capability, cycling stability, and a very promising electronic conductivity. The use of CNTF fabrics as scaffolds in composite electrodes or as current collectors, as well as prelithiation, could help exploiting these favorable properties, while also limiting the impact of the mentioned drawbacks.[3,4,24] Partial amorphization of the CNTFs is unlikely to limit electrode conductivity in full cells, but amorphization could be largely reduced through the use of more crystalline CNTFs.

## Conflict of Interest

The authors declare no competing financial interest.

## Acknowledgements

M.R. and J.J.V. are grateful for generous financial support provided by the European Union Seventh Framework Program under grant agreement 678565 (ERC-STEM) the Clean Sky Joint Undertaking 2, Horizon 2020 under Grant Agreement Number 738085 (SORCERER), by the MINECO (RyC-2014-15115) and by the Air Force Office of Scientific Research of the US (NANOYARN FA9550-18-1-7016). R.M. thanks the Spanish Ministry of Science, Innovation and Universities through the SUSBAT project (RTI2018-101049-B-I00) (MINECO/European Regional Development Fund, UE) and the Excellence Network E3TECH (CTQ2017-90659-REDT) (MINECO, Spain). N.B. acknowledges the Community of Madrid for the postdoctoral fellowship 2018-T2/AMB-12025.

## Supporting Information

Supporting information includes high resolution TEM images of pristine MWCNTF, thermogravimetric analysis of pristine MWCNTF, cyclic voltammograms, galvanostatic cycling, and Raman spectra of DWCNTF, galvanostatic cycling of MWCNTF, equivalent circuits utilized in the impedance spectroscopy analysis, impedance spectra and fitting parameters of MWCNTF.

**TOC Image**

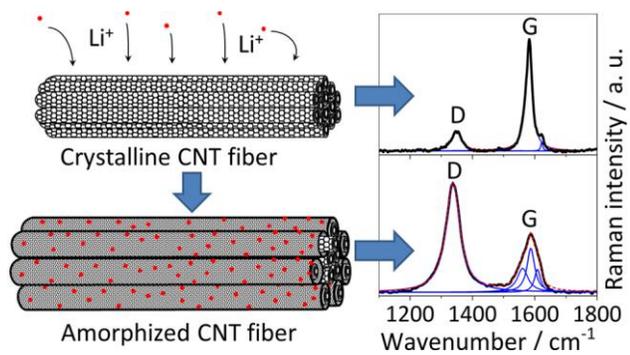